# Optical microelectrode arrays for differential readout of electrical and mechanical signals in cardiac cells

Alessandro Leronni[1], Rosalia Moreddu[2,3*]

[1]Department of Mechanical Engineering, University of Bath, Bath, UK
[2]Istituto Italiano di Tecnologia, Genoa, Italy
[3]School of Electronics and Computer Science, University of Southampton, Southampton, UK

Correspondence: rosalia.moreddu@eng.ox.ac.uk

## Abstract

Simultaneous assessment of electrical excitation and mechanical contraction is essential for understanding cardiac cell function, yet these two processes are commonly measured with separate techniques or invasively. Here, changes in cellular electrical activity modulate local charge redistribution in optical microelectrodes and are converted into fluorescence signals, while cell contraction induces membrane displacement that contributes an additional mechanical component to the optical readout. By comparing recordings obtained in beating cells with those acquired after inhibition of contraction, we separate action-potential-associated electrostatic transduction from contractility-driven membrane motion. The approach offers a label-free route to support high-throughput *in vitro* assays for cardiotoxicity screening and electromechanical sensing. Concurrently, it unfolds the physical mechanisms governing membrane-based optical devices deployed in cardiac cell bioelectronics and mechanics.

## Introduction

The function of the heart emerges from a coordinated sequence of electrical excitation and mechanical contraction at the cellular level[1]. Cardiomyocytes generate action potentials through the movement of ions across the plasma membrane, and electrical events trigger intracellular calcium signaling and activation of the contractile apparatus[2]. The resulting shortening of the cell and deformation of its membrane constitute the mechanical output of excitation-contraction coupling[3]. Because both processes are inseparably linked in physiology, methods able to resolve electrical and mechanical behavior in the same setting are of value for cardiac research, disease modeling, and drug testing. *In vitro* investigation of cardiac cells has mostly relied on separate tools for these two functions. Electrical activity has been measured using patch clamp electrophysiology[4], extracellular devices[5], microelectrode arrays[6, 7], and optical probes sensitive to membrane voltage[8]. Mechanical behavior has been probed using edge detection[9], traction force methods[10], computational tools[11], cantilevers or deformable substrates[12], and other contractility assays[13]. Although these techniques have each provided important insights, they are often implemented independently using different instrumentation and analysis workflows. This limitation is especially relevant in the study of primary cardiomyocytes and stem-cell-derived cardiac models, increasingly used to investigate arrhythmia, cardiomyopathies, maturation state, and drug responses[14, 15]. In such contexts, electrical measurements alone may fail to reveal defects in contractility, while mechanical measurements alone may not identify altered excitability[1, 3].

Microelectrode arrays are among the most widely used platforms for non-invasive interrogation of excitable cells[7]. Conventional planar microelectrodes, however, are designed to capture extracellular electrical activity, measuring ionic currents associated with membrane depolarization and repolarization, strongly influenced by the cell-electrode interface[5]. In recent years, the development of 3D microelectrodes has expanded the capabilities of bioelectronic devices[6, 16-18]. Out-of-plane electrode geometries can improve local coupling, alter electric field distribution, and enhance sensitivity to interfacial phenomena[17]. At the same time, optical interrogation strategies have opened new possibilities for transducing electrical events into optical signals[19, 20]. Optomechanical sensors have long been of interest both for diagnostics[21] and *in vitro* applications[12]. Because a contracting cell changes its shape and exerts forces on its environment, any interface in close contact with the membrane may be affected by both processes[5]. Electrical excitation modifies charge distribution, current flow, and ionic organization near the electrode[20]. Mechanical contraction alters the local spacing between the cell and the substrate.[12] At the same time, electrical activity and its relation to cell mechanics is being considered increasingly relevant across cell types[21-24], also for its potential for opening new paths in bioinspired engineering[25].

Here, we present optical microelectrode arrays integrated within a silicon nitride membrane and fluidic packaging. Fluorescent readout is used to monitor local interfacial changes attributed to contractility and electrical activity of the cardiac cells cultured on the electrodes (**Figure 1**). In drug-free conditions, when cells are spontaneously beating, the fluorescence traces include a periodic component associated with contractility and membrane displacement (**Figure 1a**). When contraction is suppressed with blebbistatin, a myosin II inhibitor widely used to uncouple motion from excitation in cardiac cultures[26], the remaining signal reflects predominantly the electrical component via electrostatic transduction (**Figure 1b**). The two signals can be decoupled (**Figure 1c**). This approach (i) provides a parallel and minimally invasive method compatible with standard microscopy, (ii) acknowledges that in cardiac cells the sensing interface is inherently influenced by motion and uses this effect to gain information about contractility, (iii) by pharmacologically suppressing contraction, it establishes a route to isolate the electrical contribution, resolving that in membrane-based optical devices[12, 20, 27] deployed with cardiac cells the signal originates from excitation as well as from contraction.

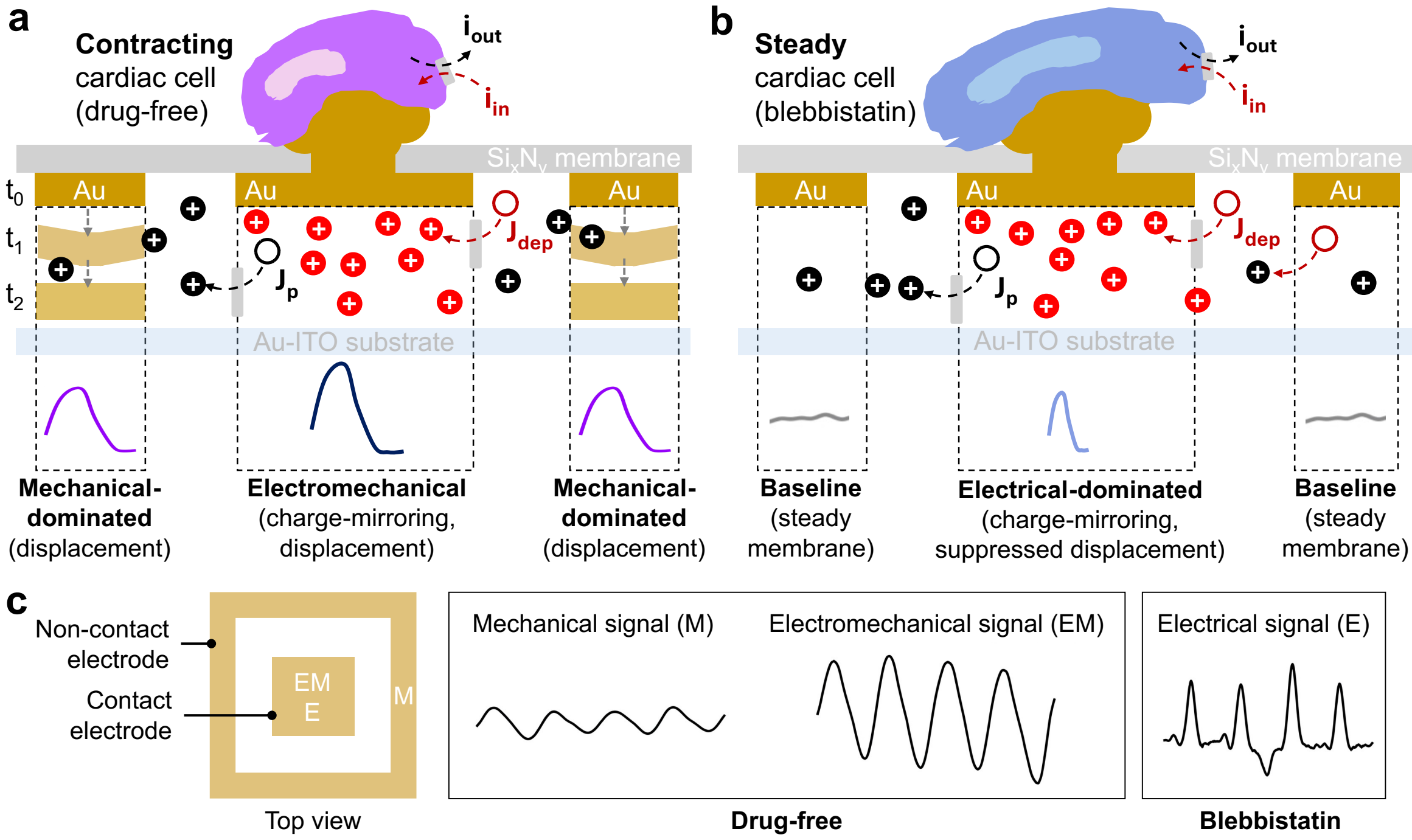

**Figure 1. Electromechanical decoupling in optical microelectrode arrays**. The schematic illustrates how the device converts cell activity into optical responses by combining a contact electrode with adjacent non-contact electrodes beneath a suspended $Si_xN_y$ membrane. Grey dashed contours labeled $t_0$ –$t_2$ indicate successive positions of the suspended membrane during a beat-associated membrane displacement event. Red circles indicate fluorophore enrichment at the contact region with consequent increase in depolarization current density $J_{dep}$, whereas black circles denote charged fluorophores in correspondence of the non-contact regions with consequent increase in polarization current density $J_p$. Currents $i_{in}$ and $i_{out}$ indicate inward and outward transmembrane ionic currents. **(a)** Under drug-free conditions, a spontaneously contracting cardiomyocyte sits over the central 3D electrode and deforms the membrane during each beat. At the lateral non-contact electrodes, this motion produces a displacement-dependent optical signal that reports a contractility-dominated output. At the contact electrode, however, two processes occur simultaneously: membrane displacement contributes a mechanical component, while action-potential-driven ionic redistribution at the electrode interface induces electrostatic transduction at the bottom interface. **(b)** Contraction is suppressed through blebbistatin treatment. Because membrane displacement is largely eliminated, the lateral non-contact electrodes no longer show a mechanics-dominated signal and are reduced to a baseline noise level. At the contact electrode, the mechanical contribution is strongly suppressed, but the electrical-induced charge redistribution remains, preserving an optical signal dominated by the action potential. **(c)** Fluorescence signal extrapolation and electromechanical decoupling from electrode pairs.

## Results

Optical microelectrode arrays were designed to generate two optically distinguishable sensing regions within each unit cell and thereby enable differential readout of cardiac electrical and mechanical activity from the same local interface (**Figure 2**). As shown in **Figure 2a**, recording units are defined by concentric square features composed of a central electrode embedded within a larger surrounding electrode frame, where the inner region corresponds to the contact electrode, which is placed in direct contact with the cell and transduces action-potential-associated interfacial charge redistribution and contractility-induced membrane displacement. The surrounding region serves as a non-contact electrode patterned on the bottom side of the membrane and senses membrane displacement associated to cell contractility. **Figure 2b** shows the array of vertical 3D contact electrodes across the contact sensing region, magnified in **Figure 2c**. 3D electrodes connecting both chambers are made possible by previously drilling holes across the membrane (**Figure S1**). Further, gold electrodeposition time controls the size of the 3D electrode base, whereas applied current tunes its surface porosity and roughness (**Figure S2**), preferred to be high in biointerfaces such to achieve tight adhesion between the cells and the electrodes[17, 20]. Vertical structures were deposited on top of the 3D electrode bases to strengthen cell-material adhesion, thereby enabling electrostatic transduction of cell signals.

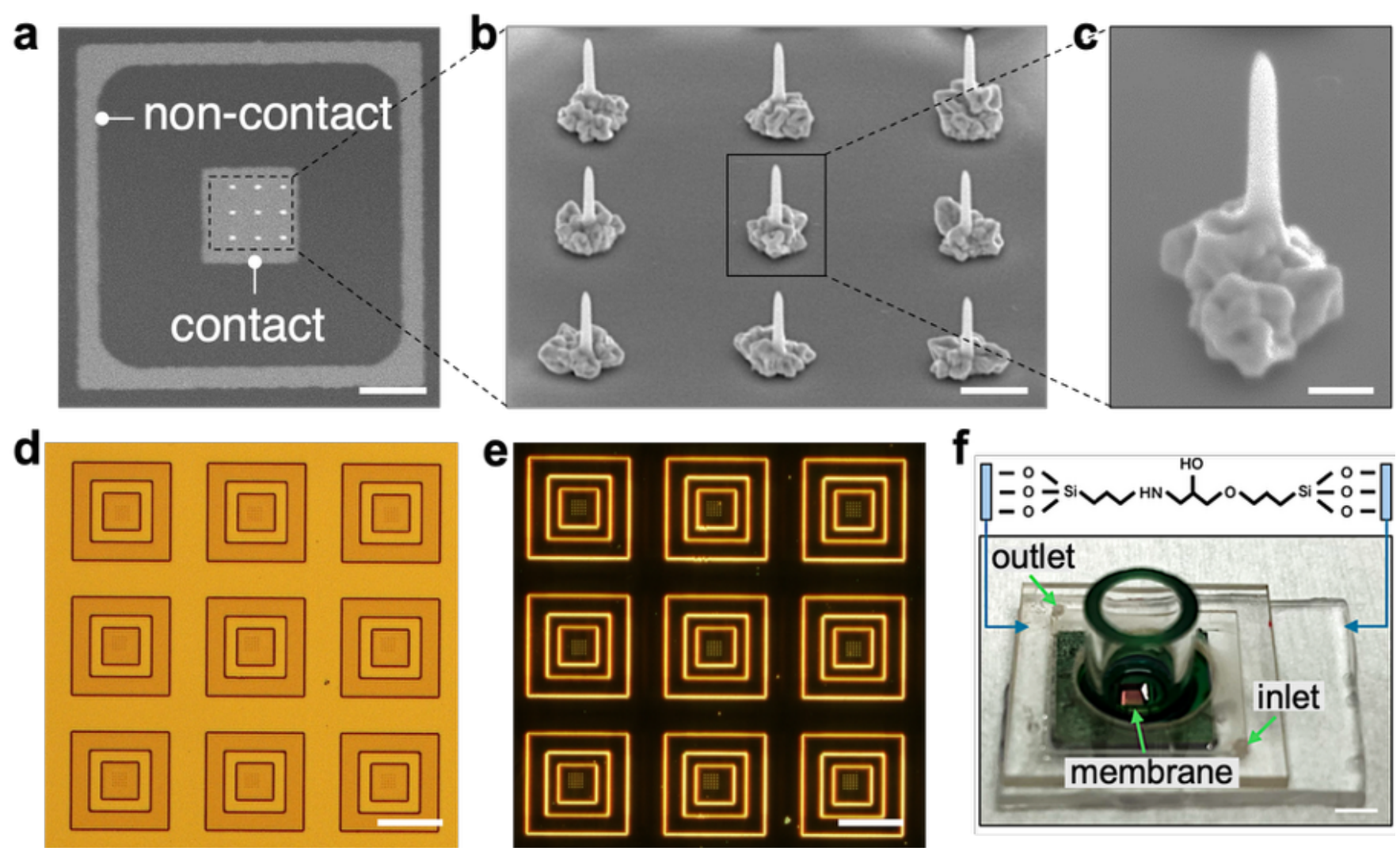

**Figure 2. Fabrication and packaging of the suspended 3D microelectrode array. (a)** SEM micrograph of a single sensing unit, showing concentric square electrode regions with a contact electrode surrounded by a larger non-contact frame. Scale bar: 20 µm. **(b)** Tilted scanning electron micrograph of a 3×3 array of 3D electrodes connecting the bottom and top of the membrane. Scale bar: 1 µm. **(c)** Higher-magnification SEM image of a 3D electrode, where the out-of-plane geometry enables tight cell interfacing. Scale bar: 300 nm. **(d)** Bright-field image of the patterned array, showing multiple sensing sites. Scale bar: 50 µm. **(e)** Dark field image of the same array, showing nanostructures in the central electrode areas. Scale bar: 50 µm. **(f)** Device packaging using a PMMA microfluidic chamber. The schematic shows the surface functionalization method used for chemical bonding[28]. Scale bar: 5 mm.

The bright-field image in **Figure 2d** displays a matrix of concentric squared electrodes for high-throughput measurements at single site resolution. The corresponding dark field image in **Figure 2e** highlights the nanostructured contact electrode areas. Such electrode pairs are patterned onto the whole membrane to achieve high throughput, with the possibility of implementation in multiple independent membranes on the same silicon chip (**Figure S3**). As shown in **Figure 2f**, the device can be packaged in a microfluidic chamber allowing to replenish the fluorophore dispersion from the inlet and the medium from the well. Measurements were performed by confocal microscopy and image analysis (**Figure 3**). **Figure 3a** shows a representative inverted-microscope image of cultured cardiac cells on the microelectrode array. Viable contracting cells are shown in **Supplementary Video S4**. **Figure 3b** shows a representative confocal microscopy frame where rhodamine fluorophores are excited at 561 nm, used for optical analysis. One electrode is highlighted to indicate the two regions of interest extracted from each electrode pair, namely the contact electrode and the surrounding non-contact electrode. The proximity of these sensing regions minimizes differences in illumination, dye environment, focal conditions, and local cellular environment. Optical recordings acquired from multiple sites across the array revealed synchronized rhythmic activity, as shown in **Figure 3c**. **Figure 3d** confirms the biological origin of the signals, where confocal measurements from cell-free electrodes exposed to medium only exhibit slow drift and low-amplitude fluctuations. This control rules out spontaneous periodicity arising from the optical system, the perfusion chamber, or the fluorescent medium.

We then examined the sensing regions under spontaneous beating conditions. The traces in **Figure 3e** show a divergence between the two optical readouts. The contact electrode trace (dark blue) shows rhythmic peaks with higher intensity. The non-contact electrode trace (purple) follows the same peak timing but remains smoother, lower in amplitude, and confined to a narrower dynamic range. This indicates that the contact electrode records a composite signal containing both action-potential-associated optical modulation and contraction-related motion effects, whereas the non-contact electrode reports predominantly the mechanical component associated with membrane displacement during contraction. This latter interpretation was verified by suppressing contraction with blebbistatin at 5 µM in cell medium (**Figure 3f**). Under this condition, the non-contact trace loses the regular oscillatory profile that characterized the spontaneously beating state in Figure 3e and instead turns into a broader and slower fluctuation that resembles the noise contribution shown earlier in Figure 3d, consistent with suppression of actomyosin-driven contraction[26] and indicating that the non-contact region functions primarily as a contractility-sensitive reporter. In contrast, the contact trace in Figure 3f retains periodic events even after contraction inhibition. Although some change in waveform relative to the drug-free condition is expected because the mechanical contribution is reduced, the persistence of recurrent signal at the contact electrode demonstrates that the residual rhythmic activity represents the electrical component arising from action-potential-induced local charge redistribution. **Figure 3g** highlights the profiles of the four signal types and shows that the largest modulation is observed at the contact electrode in drug-free medium (electromechanical signal), whereas the smallest fluctuations are measured at the non-contact electrode after contraction suppression (noise). The composite signal resulting from the contact electrode in drug-free conditions displays higher amplitude and duration compared with the same signal recorded after suppressing contractility, i.e. the electrical-dominated component. Consistent with this trend, peak-to-peak analysis across recordings in **Figure 3h** shows a reduction in signal amplitude from the drug-free contact electrode to the contact electrode signal following blebbistatin treatment and to the drug-free non-contact electrode signal.

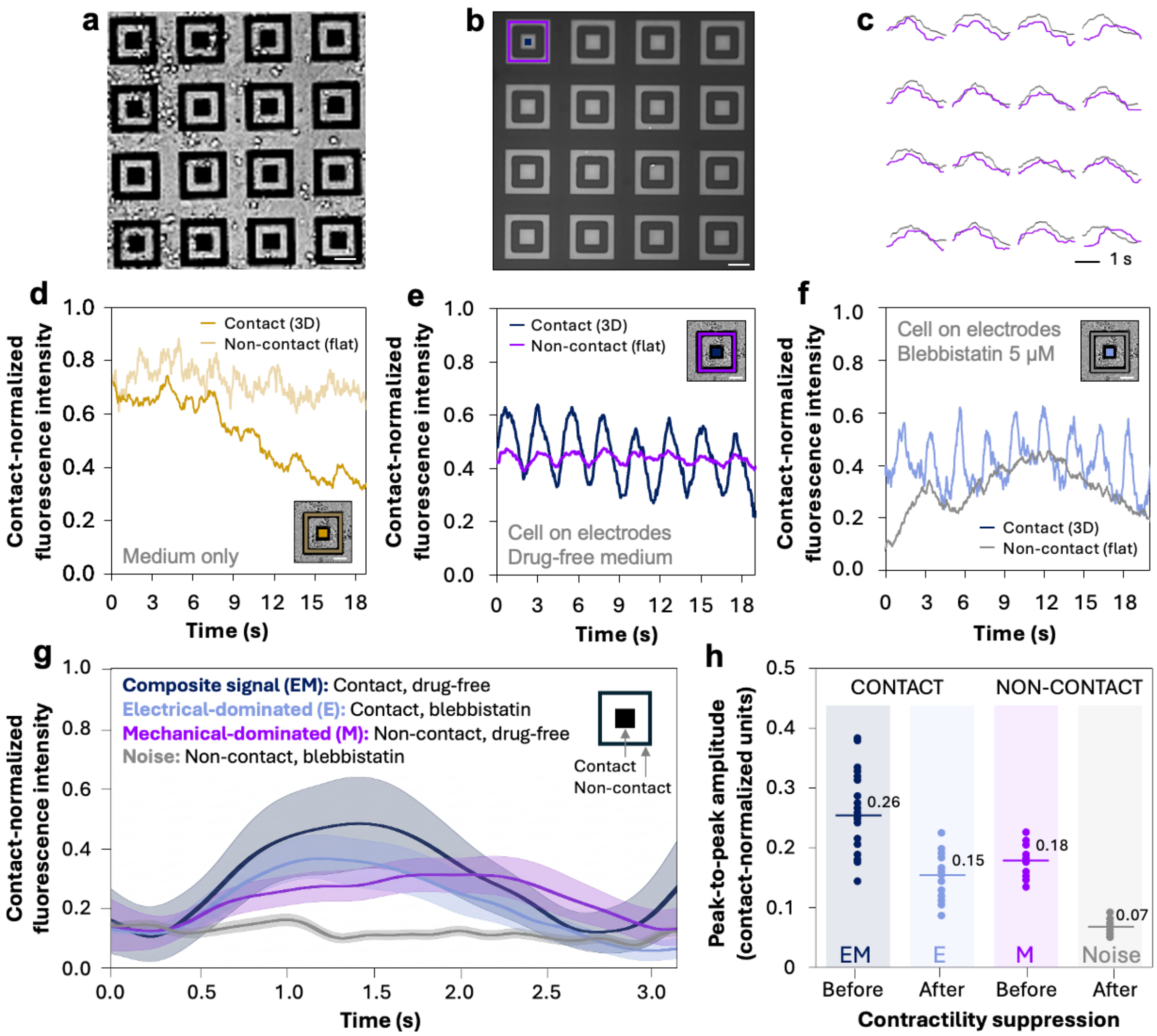

**Figure 3. Optical responses of contact and non-contact electrodes under cell-free, drug-free, and contraction-inhibited conditions.** For comparison, traces from each sensing unit are normalized to the range of the contact electrode, with the same scaling applied to the paired non-contact trace to preserve relative amplitude and timing differences between channels. **(a)** Optical micrograph of cardiomyocytes cultured on the microelectrode array. Scale bar: 30 µm. **(b)** Confocal micrograph of the microelectrode array under light excitation during optical fluorescence recording. Scale bar: 50 µm. **(c)** Time-resolved electromechanical traces recorded from the electrode pairs shown in (b). **(d)** Representative fluorescence traces recorded in medium only. Both channels exhibit low-amplitude fluctuations and gradual drift, but no periodic transients, indicating that the electrode structure, membrane, and solution do not generate signals in the absence of cells. Inset scale bar (d-e): 30 µm. **(e)** Representative paired signals from an electrode occupied by a cardiac cell in drug-free medium. The inset shows the electrode areas corresponding to the curves in d-f. Inset scale bar: 30 µm. **(f)** Representative paired signals from a cell-covered electrode after addition of contraction-inhibiting blebbistatin. **(g)** Exponentially smoothed (0.9) representative traces of the four signal classes recorded from the device, shown as mean waveform across detected peak events with the paired non-contact windows aligned to the same event times (solid lines), and shaded area displaying the standard deviation, obtained from 48 detected events across 12 electrodes in 3 devices: composite electromechanical signal (EM) at the contact electrode in drug-free medium, electrical-dominating component (E) at the contact electrode after blebbistatin, mechanical-dominating component (M) at the non-contact electrode in drug-free medium, and baseline noise at the non-contact electrode after blebbistatin. **(h)** Comparison of the peak-to-peak signal amplitudes before and after contraction suppression at contact and non-contact electrodes obtained from 12 electrodes across 3 devices.

## Discussion

The central finding of this study is that the optical signal recorded at the membrane interface contains separable contributions from action-potential-associated charge redistribution and contraction-driven membrane displacement. The paired architecture of the device, which we previously introduced in the context of non-excitable cells[20], along with a contact-electrodes-only architecture previously reported to record action potentials[27], makes this distinction accessible. To rationalize the origin of the optical signals, we developed a time-dependent Poisson-Nernst-Planck model of the fluorescent chamber, in which cationic rhodamine fluorophores and chloride counter-ions redistribute under the electric field generated by the cell and the flow induced by membrane motion (**Figure 4**, model geometry in **Figure S5**).

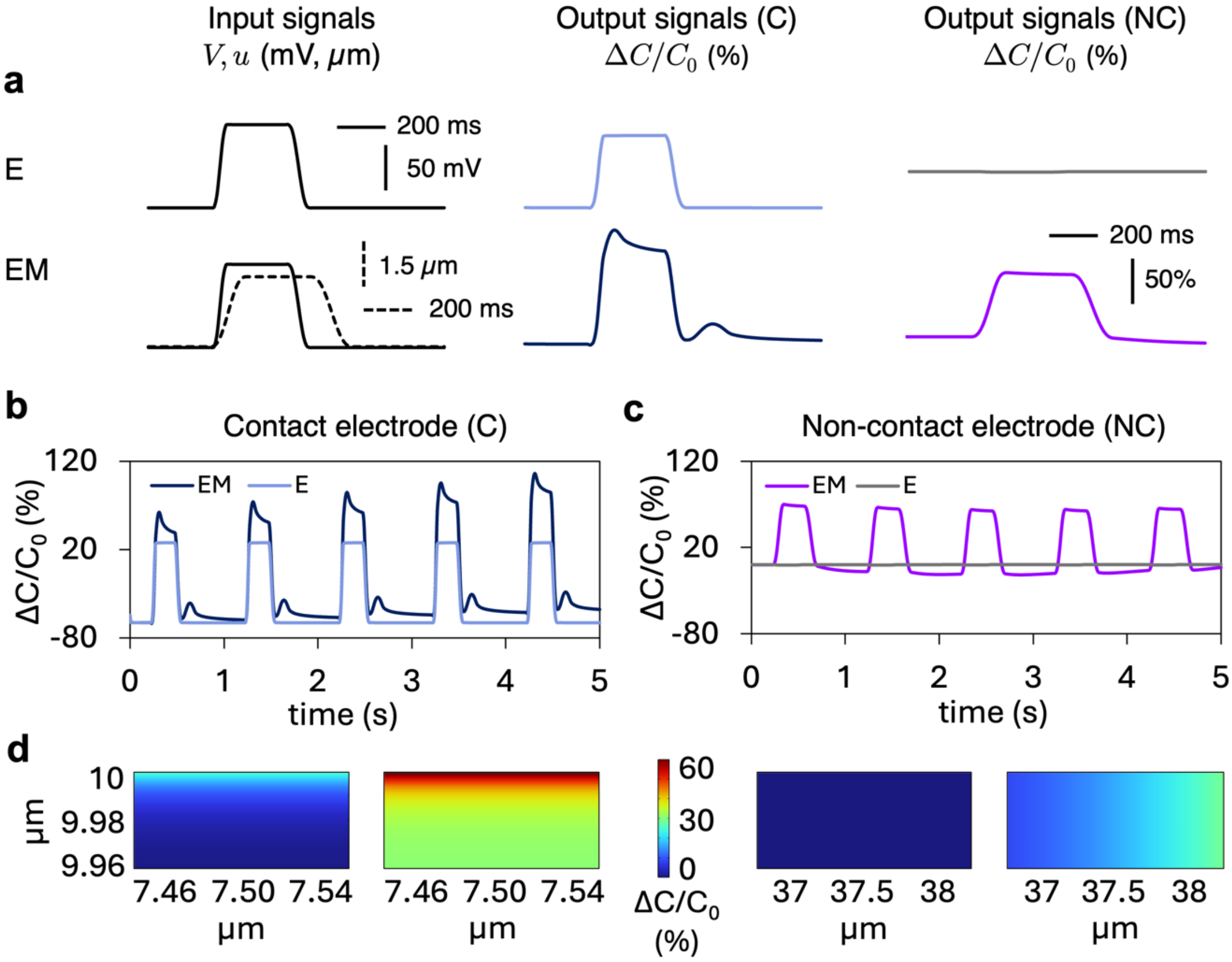

**Figure 4. Coupled Poisson-Nernst-Planck simulations of fluorophore and chloride counter-ion transport in the ethylene glycol chamber beneath the membrane. (a)** Schematic of input and output waveforms for one beat. Left: membrane potential V (solid; −85 to +20 mV, 250 ms) and membrane displacement u (dashed; 0 to 2 μm, onset 20 ms after depolarization, duration 350 ms). Center: average change in fluorophore concentration relative to the bulk value, $\Delta C/C_0$, at the contact electrode (C). Right: $\Delta C/C_0$ at the non-contact electrode (NC). **(b)** $\Delta C/C_0$ at the contact electrode over five 1 Hz cycles. The E response (light blue) switches between −63% and +27% and is identical in every cycle; the EM response (dark blue) overshoots to +62% at contraction onset, relaxes to ~40%, shows a secondary transient upon membrane return, and drifts upwards cycle after cycle (+106% in the fifth cycle) because the concentration does not recover its initial value within one beat. **(c)** $\Delta C/C_0$ at the non-contact electrode, which carries no prescribed surface charge. The E case (grey) produces no response, whereas the EM case (purple) yields a ~70% plateau that follows membrane displacement, identifying this region as a purely mechanical reporter. **(d)** Spatial maps of $\Delta C/C_0$ in the fluid layer immediately adjacent to the electrode surface (y = h = 10 μm) around the center of the contact electrode (left pair) and non-contact electrode (right pair),

for the E (left) and EM (right) cases at t = 0.40, 0.31, 0.4, 0.50 s (left to right). At the contact electrode, fluorophore enrichment is confined to a few tens of nm from the electrode, i.e. within the confocal plane.

The electrical waveform is imposed as a surface charge on the contact electrode, and contraction as a 2 μm downward membrane displacement lagging depolarization by 20 ms and outlasting it by 100 ms (**Figure 4a**). In the purely electrical case (E), depolarization reverses the electrode charge, and the fluorophore concentration at the contact electrode jumps from −63% to +27% of the bulk value, returning upon repolarization. Enrichment is confined to a nanometric layer at the electrode surface (**Figure 4d**). In the electromechanical case (EM), membrane motion convects fluorophores towards the electrode, producing an overshoot to +62%, a slow relaxation and a secondary transient when the membrane recedes. Because the mechanical contribution does not fully recover within one beat, the EM response drifts upwards over successive cycles (+106% at the fifth), whereas the E response is stationary (**Figure 4b**). At the uncharged non-contact electrode, E yields no signal while EM produces a plateau concentration change of ~70% relative to the bulk value (**Figure 4c**). Predicted peak-to-peak ranges (0.90, E-C; 1.26-1.58, EM-C; 0.77, EM-NC) scale with measured ones (0.15, E-C; 0.26, EM-C; 0.18, EM-NC, see Figure 3h) by factors of 6,4.8–6.1, and 4.3, respectively, consistent with a common transduction mechanism.

Compared with established electrophysiological methods, the present approach is not meant to replace a specific technique. Rather, it proposes a combined measurement modality. Existing techniques accurately record electrical activity alone. Patch clamp remains the benchmark for membrane voltage measurement, sampled at 10-50 kHz with sub-ms temporal resolution, but remains limited by invasiveness and throughput[29]. Conventional microelectrode arrays improve throughput[30]. Optical voltage imaging can achieve 500-1000 frames/s, and in specialized implementations exceeds 1 kHz, but it generally requires voltage-sensitive dyes or genetically encoded reporters and remains vulnerable to bleaching and phototoxicity[8]. On the mechanical side, bright-field or edge-detection contractility assays are widely used but provide displacement without electrical specificity[9]. Traction force microscopy can quantify stresses in the range of 10-$10^3$ Pa or forces in the nN regime, but it requires computational reconstruction[13]. Cantilever- and AFM-based methods offer high force sensitivity, yet their throughput is low[31] or integration with parallel electrical recording is limited[12]. In this landscape, the main advantage of the present method is the simultaneous measurement of electrical-dominated vs. mechanical-dominated signals in a single acquisition. The work has three main limitations. First, it demonstrates discrimination of signal components but does not fully recover membrane voltage or contractile force, although both have been targeted with derived methods[12, 20]. Second, the current separation strategy relies on pharmacological uncoupling with blebbistatin. Future work could separate electrical and mechanical contributions directly from waveform shape, timing, or decomposition. Third, comparison against patch clamp, MEAs, voltage imaging, and contractility assays could better define detection limits, accuracy, and broader usability of the platform. With further calibration, this strategy could support cardiac phenotyping and drug-response profiling, making it suitable for *in vitro* studies that require parallel electrophysiological and contractile recordings.

## Materials and methods

*Microfabrication.* The devices were fabricated on silicon chips incorporating a centrally suspended silicon nitride membrane, which served as the active sensing area and as a physical barrier between the upper biological chamber and the lower optical chamber in a fabrication process that we previously reported in detail[20]. To establish electrical communication across the membrane, arrays of through-membrane nanoholes with diameter of 100 nm were first defined across the membrane. Gold was then electrodeposited through nanoholes to generate contact 3D microelectrodes. Nanopillars with 1 μm height were then individually deposited in each matrix via SEM. The resulting architecture enabled direct electrical continuity between the upper and lower sides of the membrane. The lower-side gold film was patterned by standard photolithography, defining a central square electrode surrounded by a concentric hollow-square electrode. The central electrode was electrically connected to the 3D contact microelectrodes and therefore to the upper biological chamber. In contrast, the surrounding reference electrode was confined to the lower optical side and interfaced only with the fluorescent solution in the optical chamber. To enable cell culture, a glass cylinder was bonded to the top side of the silicon chip above the suspended membrane via PDMS curing. During measurements, the lower optical chamber was assembled by placing the device chip on top of a glass counter-electrode substrate coated with indium tin oxide (ITO) and gold. A droplet of rhodamine-based fluorescent dispersion was deposited on the counter electrode before assembly, thereby filling the space between the patterned planar electrodes and the lower glass substrate. In this configuration, the upper chamber contained the cells and culture medium, whereas the lower chamber contained the fluorescent optical medium. The device was assembled into a bespoke PMMA microfluidic chamber fabricated via laser micromachining.

*Cell culture.* Human induced pluripotent stem cell-derived cardiomyocytes (hiPSC-CMs; FUJIFILM Cellular Dynamics, Inc.) were used as the cardiac model and cultured according to standard protocols reported in the literature[12]. Briefly, before cell seeding, devices were cleaned via oxygen plasma to increase hydrophilicity, sterilized under UV light, and coated with an extracellular-matrix adhesion layer. Cells were thawed and seeded directly onto the active membrane area in plating medium. After cell adhesion, fresh maintenance medium was added, and cultures were maintained in an incubator at 37°C and 5% $CO_2$. Medium was replaced regularly according to standard hiPSC-CM culture practice. Recordings were performed on mature spontaneously beating cultures after 13 days *in vitro*, when synchronous contractile activity was stably observable. For contraction-inhibition experiments, blebbistatin was administered via the upper chamber immediately before imaging at a concentration of 5 μM. Medium-only control measurements were acquired on devices prepared identically but not seeded with cells.

*Confocal microscopy.* Optical recordings were acquired using a laser-scanning confocal microscope equipped with a resonant scanner (Nikon A1R) and a 20× objective at 30 fps. Rhodamine fluorescence was excited at 561 nm, and emission was collected in the 570-620 nm range. The focal plane was set on the patterned electrode layer in the optical chamber. The lower chamber contained rhodamine 6G dispersed in ethylene glycol at 0.2 mg $mL^{-1}$, whereas the upper chamber contained cultured human-derived cardiomyocytes and cell medium. Time-lapse image sequences were acquired from fields containing matrices of ~36 sensing units. For each experiment, the same acquisition settings were maintained across control, drug-free, and blebbistatin conditions to allow direct comparison of signal amplitude and waveform. Bright-field images were collected to verify membrane integrity and cell coverage of the sensing area. For each sensing unit, two regions of interest (ROIs) were defined: one corresponding to the contact electrode and one corresponding to the surrounding non-contact electrode. Because the two ROIs were concentric and closely spaced, they experienced similar illumination, fluorophore environment, and focal conditions.

*Data analysis.* Image sequences were exported as TIFF stacks and analyzed offline in Fiji/ImageJ. For each ROI, the mean fluorescence intensity was extracted frame by frame and electrode by electrode to obtain a time-dependent optical trace. Slow baseline drift caused by gradual changes in illumination, focal offset, or fluorophore redistribution was removed by subtracting the local baseline. The same workflow was applied to contact, non-contact, and cell-free control recordings. Spike-related events were identified from the contact channel. Under drug-free conditions, signals observed in both channels with matched timing were interpreted as containing a mechanical component, whereas signals that persisted selectively at the contact electrode after blebbistatin treatment were assigned predominantly to electrostatic transduction. Medium-only recordings were processed identically and were used to evaluate the absence of periodic activity. Because paired contact and non-contact ROIs exhibited a systematic baseline fluorescence offset arising from electrode geometry (**Figure S6**), traces from each sensing unit were normalized to the 0-1 dynamic range of the contact electrode, and the same scaling was applied to the paired non-contact trace before comparison in Figure 3, preserving relative amplitude and timing differences between channels and electrode types.

*Modeling.* Ion transport and electrostatics in the fluid layer of the sensor were described by a Poisson-Nernst-Planck system solved by the finite element method in COMSOL Multiphysics 6.1. The electric potential $\phi$ satisfies Poisson's equation

$$\nabla \cdot \boldsymbol{D} = F(C^{+} - C^{-}) \tag{1}$$

where $\nabla \cdot$ is the divergence operator, $\boldsymbol{D} = \varepsilon \boldsymbol{E}$ is the electric displacement, $\boldsymbol{E} = -\nabla\phi$ is the electric field (with $\nabla$ the gradient operator), $F$ is the Faraday constant, and $C^{+}$ (denoted as $C$ in Figure 4 for simplicity) and $C^{-}$ are the molar concentrations of cationic fluorophores and chloride ions, respectively. The permittivity is given by $\varepsilon = \varepsilon_0 \varepsilon_r$, with $\varepsilon_0$ vacuum permittivity and $\varepsilon_r = 37$ as the relative permittivity of ethylene glycol. The Nernst-Planck equation for the fluorophores, describing their motion, is given by

$$\frac{\partial C^{+}}{\partial t} + \nabla \cdot \boldsymbol{J}^{+} = 0 \tag{2}$$

with flux

$$\boldsymbol{J}^{+} = -D^{+}\left(\nabla C^{+} + \frac{F z^{+} C^{+}}{RT} \nabla\phi\right) + C^{+}\boldsymbol{v} \tag{3}$$

accounting for diffusion and electromigration, and the advective term for fluid motion induced by membrane displacement. In the above equation, $D^{+}$ is the fluorophore diffusivity in ethylene glycol, $z^{+} = +1$ is the fluorophore valency, $R$ is the gas constant, $T$ is the absolute temperature, and $\boldsymbol{v}$ is the fluid velocity. An analogous equation holds for chloride ions. Diffusivities were obtained from literature values in water[32] by Stokes-Einstein scaling with the water-to-ethylene-glycol viscosity ratio[33], giving $D^{+} = 1.61 \times 10^{-11}\ \mathrm{m^2 s^{-1}}$ and $D^{-} = 1.12 \times 10^{-10}\ \mathrm{m^2 s^{-1}}$ at $T = 293.15$ K. Electrodes were treated as ideally polarizable, the solution as dilute and the permittivity as constant. The computational domain is a 2D cross-section through the center of one sensing unit (**Figure S5**). Membrane motion was represented on a fixed domain by the incompressible velocity field

$$\begin{cases} v_x = \frac{v}{h}x \\ v_y = -\frac{v}{h}y \end{cases} \quad (4)$$

where $v = \frac{du}{dt}$ is the vertical membrane velocity and $u(t)$ is the vertical membrane displacement. The action potential was imposed as a surface charge density on the contact electrode

$$\rho_s = -cV(t), \quad (5)$$

with $c = 0.01\ \mathrm{F\ m^{-2}}$ the specific membrane capacitance and $V(t)$ the membrane potential, so that depolarization renders the electrode negatively charged and attracts fluorophores. All other boundaries, including the non-contact electrode, carry zero surface charge, and the gauge was fixed by constraining the domain integral of $\phi$ to zero. Zero normal ion fluxes were prescribed at the bottom substrate, $y = 0$, and at the lateral boundaries of the chamber, $x = \pm L/2$, due to symmetry with adjacent sensing units. At the top of the chamber, $y = h$, normal ion fluxes are purely advective due to membrane motion. Initial concentrations were uniform and electroneutral with $C_0 = 0.418\ \mathrm{mol\ m^{-3}}$ (0.2 mg $\mathrm{mL^{-1}}$ rhodamine 6G). $V(t)$ is periodic with period 1 s: starting at $t = 0.225$ s of each cycle it rises from −85 mV to +20 mV, remains depolarized for 250 ms and returns to −85 mV, with ~50 ms ramps. In the electrical case (E) $u = 0$, whereas in the electromechanical case (EM) the displacement $u(t)$ starts 20 ms after depolarization onset, rises to 2 µm in ~100 ms and returns to zero 350 ms after onset, outlasting the action potential by 100 ms.

## Conclusion

This work shows that the optical response of membrane-integrated microelectrode interfaces in cardiac cells contains two interpretable contributions: an electrical-dominated component associated with charge mirroring at the contact electrode and a mechanical-dominated component arising from contraction-dependent membrane displacement. By combining paired contact and non-contact regions within the same sensing unit, the platform enables local comparison of coupled electromechanical behaviors and is compatible with microscopy and existing assays. The findings provide a practical basis for experiments in which excitability and contraction need to be interrogated simultaneously and non-invasively, including pharmacological studies in stem-cell-derived cardiac models. Further calibration may allow to relate both optical traces quantitatively to membrane voltage and contractile force, opening a path toward compact multimodal platforms for cardiac phenotyping, safety assessment, and studies of excitation-contraction coupling at high-throughputs.

## Acknowledgements

AL acknowledges the University of Bath for computational resources. RM acknowledges technical support and facility access at IIT (cell culture and confocal microscopy, cleanroom) and at the University of Birmingham (laser microfabrication), the European Commission for a Marie Skłodowska-Curie Individual Fellowship “COSMOS” (grant agreement number 101064443), and the European Union’s Horizon 2020 Research and Innovation Programme

"TOX-Free" (grant agreement number 964518). RM acknowledges support from the University of Southampton through a MULSER equipment fund (2025).

## Author contributions

RM conceived the idea of using membrane-based devices to measure cardiac cell contractility and electromechanical coupling, opening this direction at IIT in 2022. RM fabricated the devices and performed the measurements. RM and AL analyzed the data and prepared the figures. AL built the multiphysics model and analyzed computational results. RM and AL wrote and revised the manuscript. RM led the project.

# Supplementary material

**Content:** Figure S1-S3, Supplementary Video S4, Figure S5-S6.

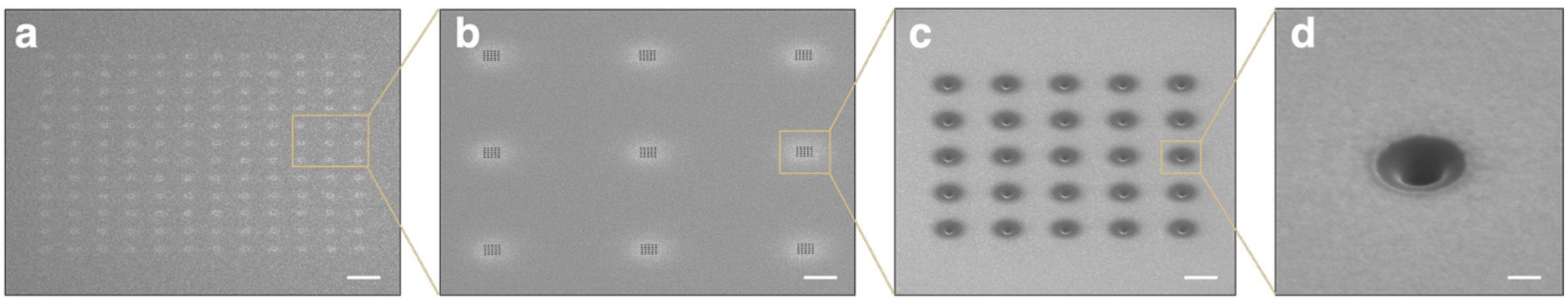

**Figure S1. SEM images of FIB-drilled membrane nanohole arrays. (a)** Suspended membrane after FIB patterning, showing the distribution of nanohole-arrays across the active area. Scale bar: 200 µm. **(b)** Selected patterned sites, each corresponding to one contact electrode sub-array. Scale bar: 60 µm. **(c)** Single nanohole array. Scale bar: 400 nm. **(d)** Representative nanohole formed during ion milling. Scale bar: 100 nm.

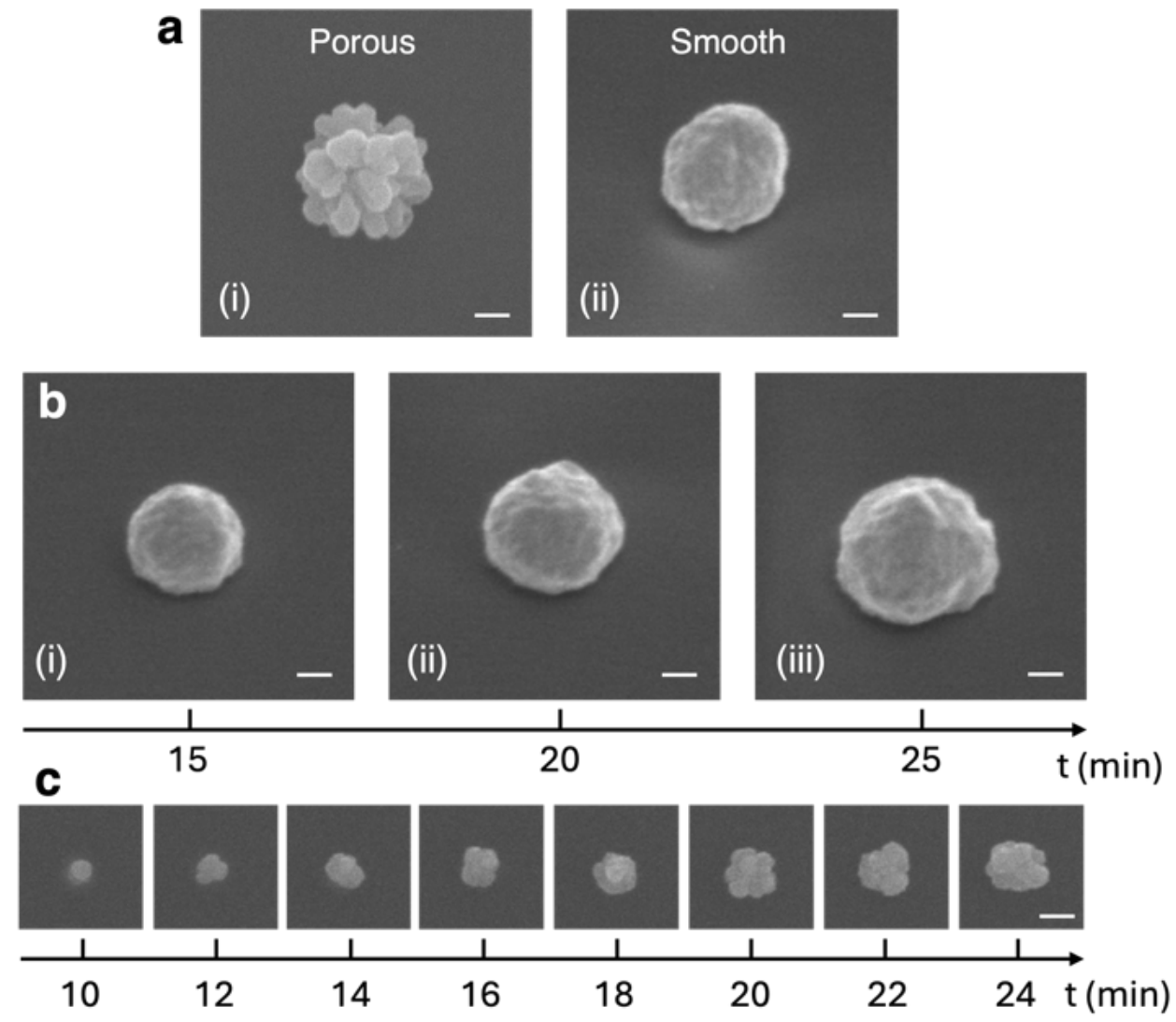

**Figure S2. SEM micrographs show the effect of electrodeposition conditions on the morphology and volume of 3D gold electrode bases grown through nano-holes. (a)** Electrodes deposited for the same total time (20 min) under two regimes: (i) high current density yields a porous structure; (ii) lower current density yields a smoother structure. Scale bars: 300 nm. **(b)** Time course of the smooth electrode at 15, 20, and 25 min. Scale bars: 250 nm. **(c)** Time course of the porous electrode from 10 to 24 min. Scale bar: 800 nm.

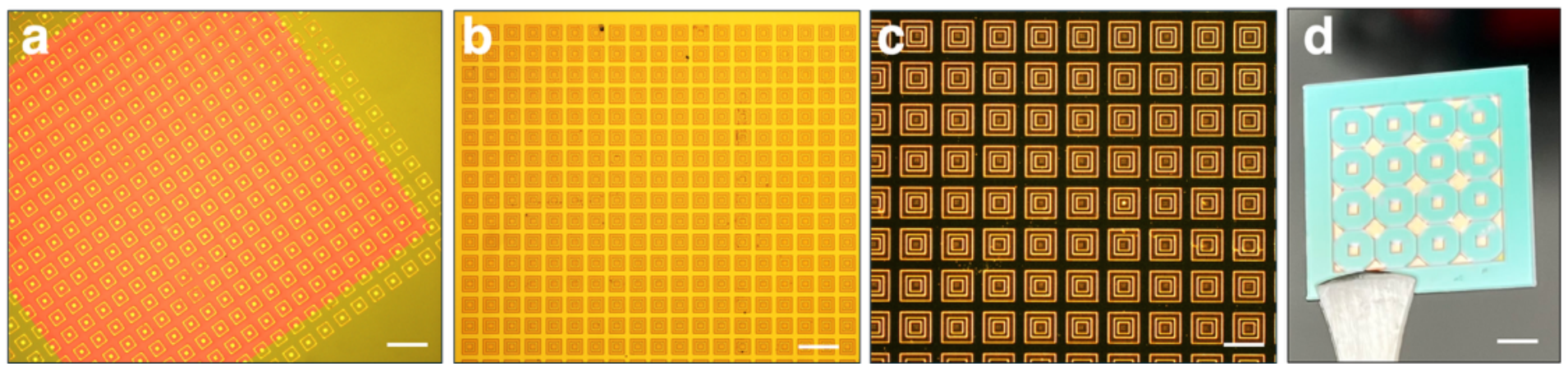

**Figure S3. High-throughput capability and prospective scaling of the technology. (a)** Optical micrograph of a membrane (orange) on silicon (olive green) patterned with electrode pairs. Scale bar: 50 µm. **(b)** Representative electrode lattice obtained by planar metallization and lithographic patterning of the membrane. Scale bar: 200 µm. **(c)** Dark-field matrix of electrode pairs, where 3D structures passing through the membrane across the contact electrodes are visible. Scale bar: 120 µm. **(d)** Integration of multiple membranes into a single chip, enabling parallelization of measurements across multiple independent recording areas on the same silicon substrate. Scale bar: 2.5 mm.

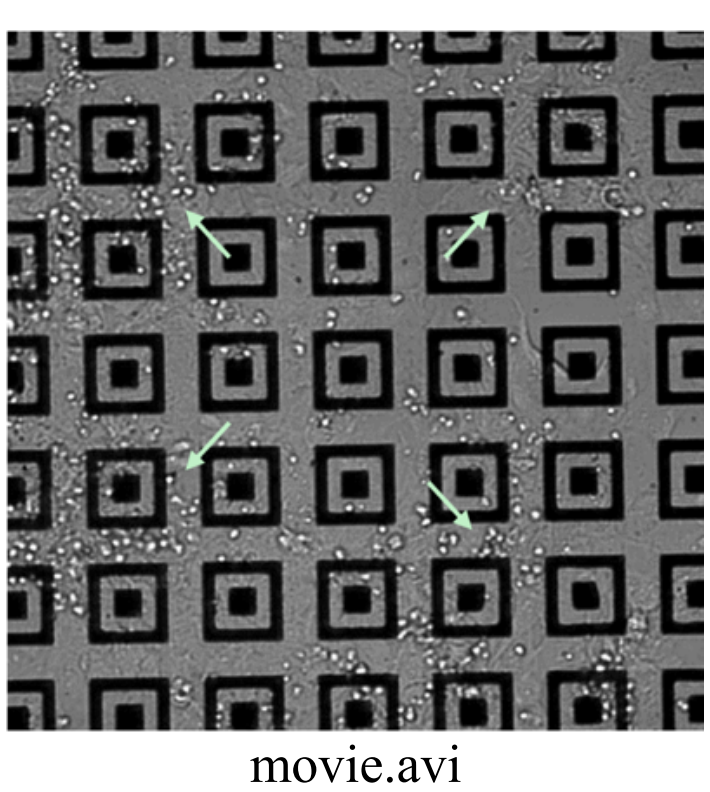

movie.avi

**Supplementary video S4.** Contracting cardiac cells cultured on the device in drug-free medium.

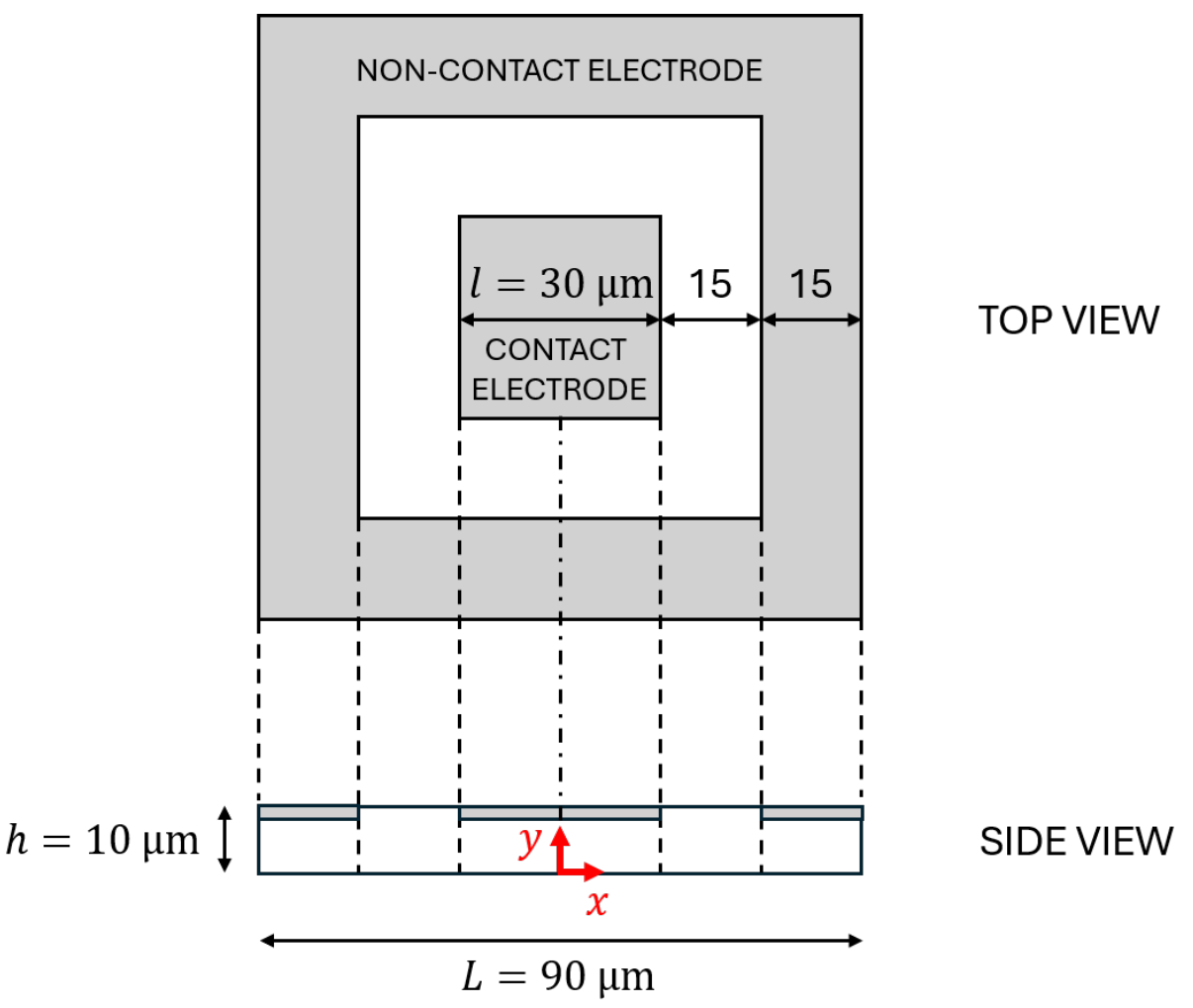

**Figure S5.** Geometry of the computational model.

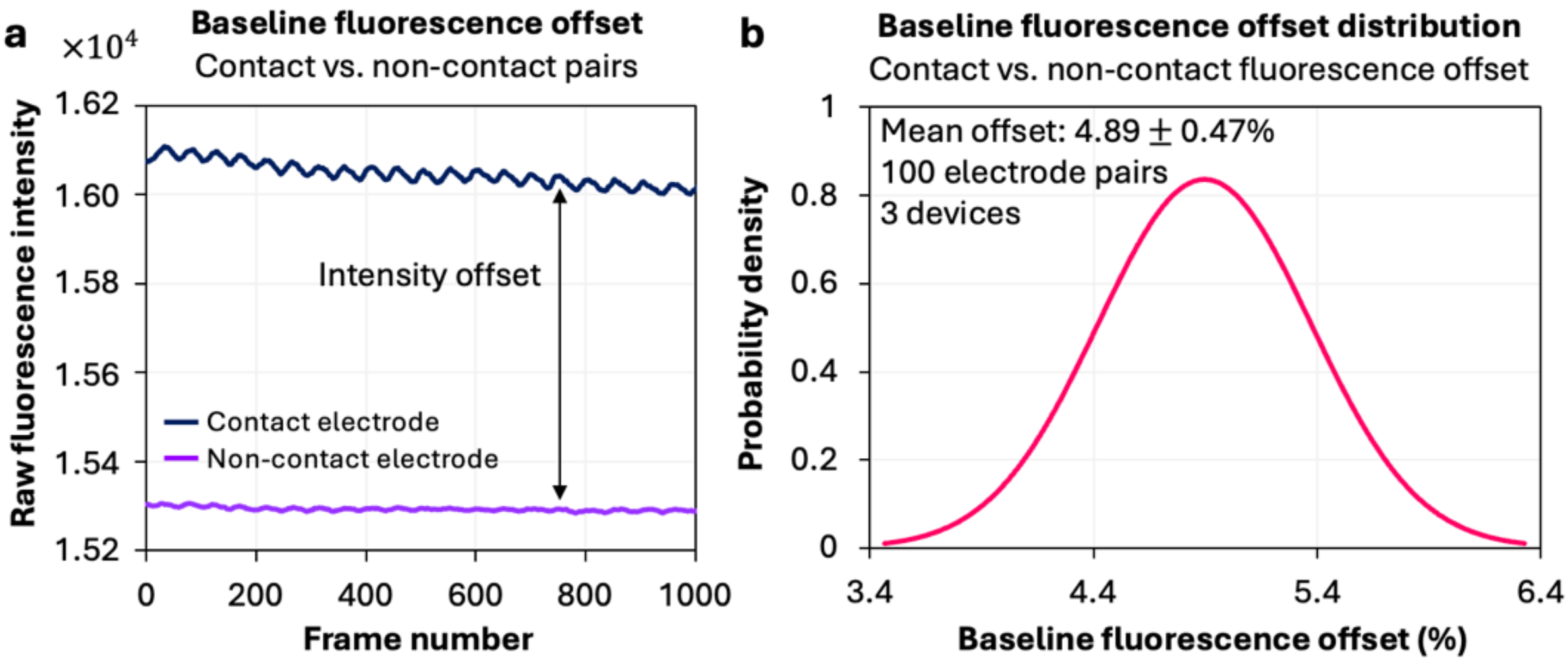

**Figure S6.** Baseline fluorescence offset between contact and non-contact ROIs. **(a)** Representative raw fluorescence traces from an electrode pair before normalization, showing a systematic baseline fluorescence intensity offset between the contact and non-contact electrodes. **(b)** Distribution of the baseline fluorescence offset between paired contact and non-contact ROIs across 100 sensing units from 3 devices. The mean baseline difference was 4.89 ± 0.47% (mean ± SD). Baseline fluorescence offset was calculated as $100 \times (F_{contact} - F_{non-contact}/F_{contact})$. This offset motivated the normalization applied to the time-resolved traces in Figure 3.